\documentstyle[aps,prl,epsfig,graphics]{revtex}

\begin{document}

\draft

\twocolumn[\hsize\textwidth\columnwidth\hsize\csname
@twocolumnfalse\endcsname

\title{Quantum Measurement and Shannon Information, A Reply to M. J. W. Hall}
\author{\v Caslav Brukner and Anton Zeilinger}
\address{Institute for Experimentalphysics, University of Vienna,\\
         Boltzmanngasse 5, A--1090 Vienna, Austria}

\date{\today}

\maketitle

\vskip1pc]

\begin{abstract}

Motivated by Hall's recent comment in quant-ph/0007116 we point
out in some detail the essence of our reasoning why we believe
that Shannon's information is not an adequate choice when defining
the information gain in quantum measurements as opposed to
classical observations.

\end{abstract}


In a recent comment \cite{hall}, Hall raises some very interesting
points with respect to our recently published criticism
\cite{brukner} of the applicability of Shannon's measure
\cite{shannon} of information  to quantum measurement. Before
answering some of Hall's criticism directly we would like to
recall some of the essential points of our view:

1) We require from an information measure appropriate to a quantum
experiment that it firstly describes the information gain of an
individual experiment and secondly that the total information gain
is described by the sum of a complete set of non-commuting, i.e.
complementary, observables.

2) We require that the information gain be directly based on the
observed probabilities, (and not, for example, on the precise
sequence of individual outcomes observed on which Shannon's
measure of information is based).

3) As the particular choice of the complete set of non-commuting
observables is at the whim of the experimentalist, and as the
total information content of a quantum system must be independent
of the experimentalist's choice, we require the total information
content to be invariant upon that choice.

For the purpose of further discussion we denote the measure of
information in the measurement of observable $\hat{A}$ by
$I(\vec{p})$ where $\vec{p}=p_1,...,p_n$ are probabilities for
outcomes to occur. Then the total information content of the
quantum system is defined as the sum of individual measures of
information over a complete set $\hat{A}_1,...,\hat{A}_j,...$ of
mutually complementary observables, i.e. $I_{total} \equiv \sum_j
I(\vec{p}_j)$ where $\vec{p}_j$ is the probability distribution
observed in the measurement of observable $\hat{A}_j$. With the
specific measure of information $I(\vec{p})$ (proportional to the
sum of the squares of probabilities) that both mathematically and
conceptually differs from the Shannon measure the total
information of the quantum system is shown to have the property of
invariance required in 3) above \cite{brukner,caslav}.

An essential point of our argument is that Shannon's measure of
information is intimately tied to the notion of systems carrying
properties prior to and independent of observation. This is
certainly the case for classical systems but, in general, this is
not true for quantum systems. It is only true for quantum systems
in the rare case of measurements performed in a basis where the
density matrix of the system is purely diagonal. Be it a pure
state, which is measured using an apparatus to whom the state is
an eigenstate, or be it a mixed state, in its diagonal
representation where the probabilities for the subsystems occur
can be viewed as classical probabilities for the measurement
results in that diagonal basis.

While Hall seems to agree that Shannon's reasoning of breaking
down a decision tree of various ways cannot be applied to
consecutive quantum measurements for non-commuting observables, he
suggests that this observation of ours is not relevant in judging
whether Shannon's information is applicable to quantum
measurement. We note that at least one of the reasons given to
justify the use of Shannon's information in general is therefore
unapplicable, yet we would suggest that we cannot agree with
Hall's point where he criticizes our use of the Fadeev
\cite{faddeev} form of the so-called grouping axiom. In particular
he suggests that the grouping axiom which, we agree, directly
leads to Shannon's information can still be used if one considers
non-overlapping distributions which are defined such that there
exists some measurement which is able to discriminate between them
with certainty.

We suggest that this is exactly the case when we observe a density
matrix in its eigenbasis where it has no non-diagonal terms, that
is, when we can indeed introduce classical probabilities. Yet,
clearly, if we consider other observables, where the density
matrix is not non-diagonal, we certainly arrive at states which
are coherent superposition of the basis states and therefore if we
perform the measurement in this new basis the distributions cannot
uniquely be defined. It should also be noted that the results
observed in one single quantum measurement can always be assumed
as emerging from a classical mixture of non-overlapping
distributions with weights being equal to the probabilities for
the results observed. Yet again, these distributions cannot be
defined uniquely for all possible non-commuting measurements
\cite{kochen}. Therefore, we suggest that this argument by Hall
actually can be viewed as supporting our general proposition.

In his comment Hall analyzes some geometrical properties of
Shannon and von Neumann entropies and he gives a connection
between them pointing to the existence of a unique measure of
uncertainty for classical and quantum systems with the geometrical
properties of a ''volume'' \cite{hall,hall1}. This leads him to
the conclusion that the von Neumann entropy is in fact an
appropriate quantum generalization of Shannon entropy. While it
might be of theoretical importance to request for an adequate
measure of information in classical and quantum physics certain
well-defined geometrical properties, we suggest to judge the
physical significance of a measure of information by its
operational properties because after all every information about
the system is obtained at only by observation.

Following this requirement we argue that, with the only exception
for results of measurement in a basis decomposing the density
matrix into a classical mixture when it is equivalent to Shannon's
information and therefore indeed refers to the information gain in
an individual measurement experiment, the von Neumann entropy is
just a measure of the purity of the given density matrix without
any explicit reference to information contained in individual
measurements.

Hall argued that this kind of criticism also holds for our
proposed total information content because for the measurement in
the basis in which the density matrix is diagonal the measure of
information $I(\vec{p})$ is equal to the total information
$I_{total}(\hat{\rho})$. We emphasize that for a complete set of
mutually complementary observables, thus including also those for
which the density matrix cannot be decomposed into a classical
mixture, and furthermore without knowledge of the basis in which
the density matrix is diagonal, we always have
$I_{total}(\hat{\rho})=\sum_j I(\vec{p}_j)$. In contrary there is
no such relation for the von Neumann entropy and the set of
individual Shannon's measures associated to mutually complementary
measurements. The case of $I(\vec{p}) = I_{total}(\hat{\rho})$ is
just an extreme one of maximal knowledge of one observable at the
expense of complete ignorance of complementary ones, however the
value of the total information remains unchanged also for other
choices of a complete set of observables where we have only
partial knowledge of them.

In relation to Hall's argument mentioned above, it might be of
interest that besides the discussed case of the measurement in the
basis in which the density matrix is diagonal there exists an
infinitely large set of individual measurements for all of which
the measure of information is indeed equal to the total
information content. We explain this on the simple example of a
spin-1/2 particle for which we know to have three mutually
complementary measurements. These are the measurements of spin
along orthogonal directions $x^1$, $x^2$ and $x^3$. From the three
von Neumann measurements we may formally constitute one single
measurement with 6 outcomes described by a POVM
\begin{eqnarray}
\frac{1}{3}\sum_j^3 |{ \bf x}^j \rangle \langle { \bf x}^j | + |{-
\bf x}^j \rangle \langle {- \bf x}^j | = \hat 1. \label{id2}
\end{eqnarray}
This is just the properly normalized sum of three completeness
relations. From this one single POVM we obtain 6 probabilities.
Inserting them in the expression for our measure of information we
get a quantity which is invariant under unitary transformation. We
find it intriguing that just from one single POVM experiment we
can obtain the total information of the quantum system which
comprises in itself all possible mutually complementary classes of
information.

Hall also suggests that the von Neumann can be viewed as an
operational concept as it is based on probabilities. We remark
that the usual procedure to obtain the von Neumann entropy
includes performance of the sufficient number of measurements
necessary for complete specification of the density matrix,
calculation of its elements using a very specific and well-defined
prescription from quantum theory that connects probabilities
observed in the measurements and complex elements of the density
matrix, and finally performing a particular mathematical operation
In the formulation of our information content we were lead by an
operationally based plausible idea that in quantum mechanics an
experimentalist can have only partial knowledge about mutually
exclusive measurements in general without any further reference to
the structure of the theory. By simple summarizing these partial
knowledges we obtain immediately a quantity which turns out to be
invariant under unitary transformation in contrast to the
procedure mentioned above where one has to use concepts from
quantum theory from the beginning.

Finally Hall makes the interesting suggestion of defining our new
measure of information not only via complete sets of mutually
complementary observables but by an average over all
non-degenerate Hermitian observables. This apparently is a nice
generalization independent of the question whether or not a
complete set of mutually complementary observables exists for
Hilbert spaces of any dimension.

We would like to thank Zdenek Hradil for helpful comments and
discussions. This work has been supported by the Austrian Science
Foundation FWF, Project No. F1506 and the US National Science
Foundation NSF Grant No. PHY 97-22614.

\narrowtext

\end{document}